# Ionic conductivity enhancements and low temperature synthesis of $Li_7La_3Zr_2O_{12}$ garnets by Bi aliovalent substitutions.

Derek K. Schwanz[1], Andres Villa[1], Mahalingam Balasubramanian[2], Benjamin Helfrecht[1], Ernesto E. Marinero[1]

[1]School of Materials Engineering, Purdue University, West Lafayette, Indiana 47907, USA.

[2]Advanced Photon Source, Argonne National Laboratory, Argonne, Illinois 60439, USA.

**ABSTRACT**

We report on a novel approach to synthesize cubic-phase fast ionic conducting garnet-type solid state electrolytes based on Bi doped $Li_7La_3Zr_2O_{12}$ (LLZO). Bi aliovalent substitution into LLZO utilizing the Pechini processing method is successfully employed to synthesize $Li_{7-x}La_3Zr_{2-x}Bi_xO_{12}$ compounds. Ionic conductivities up to $2.0 \times 10^{-4}$ S/cm are achieved in structures not fully densified. Cubic phase $Li_6La_3ZrBiO_{12}$ powders are generated in the temperature range from 650 °C to 900 °C in air. In contrast, in the absence of Bi and under identical synthesis conditions, the cubic garnet phase of $Li_7La_3Zr_2O_{12}$ is not formed below 700 °C while a transformation to the tetragonal phase is observed at 900 °C for the un-doped compound. The critical role of Bi in lowering the formation temperature of the garnet cubic phase and the improvements in ionic conductivity is investigated in this work through microstructural studies and AC impedance measurements. We ascribe the effect of Bi doping in achieving these remarkable improvements to significant enhancements at lower temperatures in the kinetics of the solid-state reaction resulting in explosive grain growth and densification of the garnet. Moreover, XAS is utilized to identify the specific atomic site where Bi is incorporated in the LLZO garnet crystalline structure.

# MAIN TEXT

Conventional liquid electrolyte-salt combinations in lithium batteries present inherent safety concerns due to dendritic growth and thermal runaway issues[1], [2]. Solid state electrolytes provide increased functionality to the cell in terms of enhanced stability, cyclability and safety [3]–[5]. However, ionic transport through solid electrolyte materials is in general orders of magnitude slower than in liquid electrolytes[6]. If the ionic conductivity and the synthesis of lithium-ion conducting solid materials can be optimized, battery safety, lifetime and performance can all be improved significantly.

$Li_7La_3Zr_2O_{12}$ (LLZO) garnet-type oxides have been shown to be promising materials for electrolyte applications on account of their relatively high ionic conductivity and good chemical stability [7]–[10]. Room temperature LLZO exists as two high-temperature stabilized polymorphs: an ordered, low ionic conductivity, $I4_1/acdZ$ tetragonal phase and a more disordered, high ionic conductivity, Ia-3d cubic phase[11]–[14]. Optimization of the $Li^+$ site occupancy and lattice parameter modifications of cubic LLZO have been attributed to improvements of ionic conductivity [15]–[18]. However, elevated high temperature heat treatments are required to achieve the cubic phase stabilization and densification required for electrolyte applications[7]. Nevertheless, site specific aliovalent dopants have proven useful for engineering LLZO material properties and its stoichiometry, allowing for lower temperature stabilization of the garnet cubic phase[19]. Dopants have also been used to modify the amount of $Li^+$ in LLZO through substitution of higher valence species onto the 24c and 16a sites for $La^{3+}$ and $Zr^{4+}$ ions respectively[9], [19]–[26]. These aliovalent dopants modify also the geometry of the $Li^+$ conduction channels[16], [25], [26]. Such studies confirm that there is an optimized $Li^+$ occupancy ratio providing the highest ionic conductivity[27]–[29]. The cubic garnet lattice has a maximum of 7.5 $Li^+$ sites per formula unit, and an optimized ionic conductivity has been determined to lie near a $Li^+$ stoichiometry of 6.5 [30]. Thus, dopants simultaneously modify the lattice spacing and the stoichiometry of the garnet species, greatly affecting the lithium-ion mobility throughout the structure.

LLZO has been typically fabricated utilizing ball milling of oxide precursors. Without dopants, it can require as much as 36 hours at 1230 °C for stabilization of the cubic phase and subsequent densification to achieve ionic conductivities on the order of $10^{-4}$ - $10^{-3}$ S/cm[7]. In contrast, sol-gel synthesis techniques such as the Pechini method have been utilized to create a more homogeneous mixture of precursor materials, reducing the activation barrier for complete mixing[22], [31], [32]. An ideal dopant should simultaneously decrease the activation barrier for compound formation while aiding in the densification. Thereby, decreasing the temperature required for cubic phase formation. Therefore, judicious dopant selection and site-specific substitution, can be employed to tailor the garnet composition for optimized ionic conductivity

Here we investigate Bi aliovalent doping on the 16c Zr site in $Li_{7-x}La_3Zr_{2-x}Bi_xO_{12}$ garnet oxides. Bi has been shown to be useful in both solid state and solution based fabrication of garnet $Li^+$ conductors due to its large ionic radius and $Bi^{5+}$ valence[31], [33]. Garnet oxides of the nominal composition $Li_{7-x}La_3Zr_{2-x}Bi_xO_{12}$ were fabricated from nitrate precursors by the citrate-gel Pechini method using a polymerized complex intermediary [Supplementary information Fig. S1]. To study dopant effects on phase evolution, polymerized complexes for the compositions $Li_{7-x}La_3Zr_{2-x}Bi_xO_{12}$ (x = 0 - 1.0) were calcined between 600 °C and 700 °C for 10 hours in an MgO crucible [34], [35].

XRD was utilized to determine the influence of Bi on the formation of the cubic phase and the phase stability as a function of garnet composition. Figures 1a and 1b show diffraction patterns for $Li_7La_3Zr_2O_{12}$ and $Li_6La_3ZrBiO_{12}$ respectively of powders calcined at 600 °C, 650 °C, and 700 °C. In agreement with similar studies [27], dopant-free LLZO converts from the $La_2Zr_2O_7$ pyrochore-type phase into an Ia-3d garnet cubic phase at 700 °C (Fig. 1a). In comparison, the Bi-doped sample (Fig. 1b) transforms into the garnet cubic phase between 600 °C and 650 °C. This doped sample forms a mixture of $La_2Zr_2O_7$ along with R-3mH $BiLa_2O_{4.5}$ when heat-treated at 600 °C and readily converts to the cubic garnet phase at 650 °C, a lower temperature than for LLZO. This only forms the $La_2Zr_2O_7$ precursor phase at 650 °C as seen in Fig. 1a. This temperature reduction for cubic phase formation indicates that Bi-doping, reduces the activation energy required. The Bi-doped cubic garnet has a lower $Li^+$ site occupancy than un-doped LLZO. It can be expected to require less thermal activation to achieve the ordering for the overall garnet structure. It is noted that in the work by Gao *et. al.* on $Li_5La_3Bi_2O_{12}$ garnets, a transition to the cubic phase is reported between 600 °C and 650 °C [31]. The diffraction pattern for the $Li_6La_3ZrBiO_{12}$ sample calcined at 650 °C indicates almost complete conversion into the cubic phase with only residual amounts of the precursor phases. In contrast, there is no evidence for formation of the cubic phase in $Li_7La_3Zr_2O_{12}$ below 700 °C.

Extended X-ray Absorption Fine Structure (EXAFS) was employed to investigate the site occupancy of bismuth in the garnet lattice. EXAFS measurements were carried out at the $L_3$-edges of bismuth and lanthanum and the K-edge of zirconium at beamline 20-BM at the Advanced Photon Source (Supplemental Fig. S2).

Representative raw EXAFS spectra are shown in Fig. 2a. The structural parameters of the dominant pair correlations, extracted by fits to the data at the various edges, are provided in Table II. For $Li_6La_3ZrBiO_{12}$, the Bi local environment is consistent with Bi atoms occupying Zr-type sites. Bismuth is surrounded by ~ 6 oxygen and ~ 6 lanthanum atoms at 2.097 ± 0.007 Å and 3.656 ± 0.014 Å. The data can be fit remarkably well by considering only occupancy of Zr-type sites; this is illustrated in Fig. 2b. There is no clear evidence for the presence of longer Bi-O or Bi-Zr/Bi correlations at distances expected for Bi occupying La-type sites with a 3+ valence. We therefore infer that Bi additions occupy the Zr-type sites exclusively, to within the accuracy of the measurements. The structural parameters, reported in Table II, reveal local distortions around the Zr and La sites to accommodate bismuth additions. Furthermore, there is an increase in the mean-squared relative displacement for many correlations, pointing to an increase in static disorder.

To evaluate Bi stoichiometric effects on microstructure and ionic conductivity, pellets were fabricated with compositions $Li_{7-x}La_3Zr_{2-x}Bi_xO_{12}$ (x = 0, 0.25, 0.5, 0.75 and 1.0) from precursor powders [Supplemental Fig. S3]. Pellet fractured surfaces were examined with scanning electron microscopy (SEM, FEI XL40) to compare their microstructure [Fig. 3a-d]. The pure LLZO pellet [Fig. 3a] exhibits sub-micron sized particles and almost no inter-particle coalescence, indicating very little sintering occurring at 900 °C. In contrast, Figs. 3b-3d show significant grain growth and evidence of sintering. The sample with the highest amount of Bi (Fig. 3d) shows the largest grain growth and a high inter-particle coalescing relative to Figs. 3b and 3c, this is indicative of a high degree of sintering at 900 °C. Therefore, when Bi is included in the garnet structure, particle growth and sintering is significantly enhanced in LLZO-doped samples. Thus, when incorporated into the synthesis of LLZO cubic garnet, Bi acts as a sintering aid and enabler of rapid grain growth; furthermore, the amount of grain growth and densification is correlated to the amount of Bi added.

Bi additions to LLZO have a remarkable impact on grain growth and sintering as evidenced in the results of Fig. 3. Such explosive grain growth could be accounted for by the formation under our experimental annealing conditions of a very low viscosity phase, resulting in greatly increased diffusivity. We note that the phase $Bi_2O_3$ with a melting point of ~ 830 °C, could lead to the formation of low melting point eutectics in combination with oxides such as $ZrO_2$[36]. Therefore, it is plausible that the formation of low melting point eutectic phases under our experimental conditions account for the explosive grain growth here reported. Similar results have been reported by Wang and Sakamoto [37]. In contrast, in the work of Wagner et al[38], no significant grain growth with increasing Bi-doping is observed. A different synthesis method and carbonate precursors are employed in their work.

We further characterized our pellets using AC impedance spectroscopy from 0.0001 Hz - 300 kHz employing a Solartron 1260 impedance analyzer. Complex impedance plots are provided in Fig. 3e and Table I summarizes measurements for the LLZO samples with various Bi content.

Complex impedance plots for pellets with different Bi amounts are displayed in Fig. 3e. Measurements were taken at 27 °C and results analyzed using equivalent circuits models as described by Huggins[39]. The tail on the low frequency spectrum indicates the capacitive nature of the ion-blocking electrodes, while the semicircles at the high frequency end correspond to the resistive-capacitive response of the bulk material. The semicircles at high frequencies are somewhat compressed, indicating separate contributions from both the bulk and grain boundaries, but there is no significant separation. As there is no clear separation between the bulk and grain boundaries, the more resistive nature of the bulk dominates the behavior in comparison to that of the grain boundaries. The compressed semicircles allow for the measurement of the total resistance, as extracted from the low frequency intercept of the Z'($\Omega$) axis, as done in similar studies reported in [14]. Measurements of the $Li_7La_3Zr_2O_{12}$ sample did not provide reliable data to calculate its ionic conductivity. This is likely due to the lack of interconnectivity and low densification that ensues at 900 °C in this sample as indicated in Fig. 3a. The Bi containing samples yield ionic conductivity ranging from $10^{-6} – 10^{-4}$ S/cm and the results are summarized in Table 1 along with other pellet physical properties. Reported impedance measurements for LLZO typically employ either sintering temperatures in excess of 1000 °C or sintering aids such as Al –intentionally added, or in the form of a impurities from the crucible [7], [13], [22], [23]. We note that Xia et al[40] investigated mixtures of $Al_2O_3$ and $Li_{6.8}La_3Zr_{1.8}Bi_{0.2}O_{12}$ employing sintering temperatures up to 1100 °C. However, the maximum value of the ionic conductivity attained in their work was 6.3 x$10^{-5}$ S/cm.

The trend of higher ionic conductivity with increasing amounts of Bi can be partially ascribed to increments in densification and grain growth associated with higher Bi concentrations as evidence in Fig. 3. Moreover, with higher concentrations of Bi, their relative density augments, derived from Archimedes measurements, as summarized in Table 1. Finally, the increased inter-particle connectivity and coalescence observed in Fig. 3 results in a lower overall sample resistance to ionic transport through the densified structure. Thus, samples containing lower amounts of Bi such as $Li_{6.75}La_3Zr_{1.75}Bi_{0.25}O_{12}$ (x = 0.25) can be expected to have a higher resistance due to larger free volume in the structure.

As the Bi stoichiometry is varied, the $Li^+$ occupancy to vacancy ratio proportionally changes, with each $Bi^{5+}$ creating an additional $Li^+$ vacancy. Thus, it can be expected that each pellet composition studied, exhibits a different $Li^+$ sub-lattice occupancy ratio, resulting in electronic structure changes of the $Li_{7-x}La_3Zr_{2-x}Bi_xO_{12}$ garnets. The $Li_{6.25}La_3Zr_{1.25}Bi_{0.75}O_{12}$ exhibits the highest measured ionic conductivity. This sample corresponds to a Li stoichiometry of 6.25, this is slightly lower than that reported value of 6.5 by Zeier [30]. The difference is likely due to microstructural and densification changes in our samples with varying amounts of Bi. To circumvent structural from electronic effects, hot pressing methods are employed to compare samples with comparable densification characteristics [15], [23]. In this work, significant microstructural and densification dependence on Bi composition is observed, therefore, both the densification and $Li^+$ occupancy may jointly contribute to changes in the ionic conductivity values reported. In the case of $Li_{6.75}La_3Zr_{1.75}Bi_{0.25}O_{12}$ samples sintered at 900 °C for 10 hours, there is insufficient thermal activation for densification, hence, even if the electronic structure for this composition were to be an optimum for ionic transport, the degree of densification is insufficient to provide an interconnected structure for optimum ionic transport. It is quite plausible that the maximum ionic conductivity reported in the samples here reported, under the sintering conditions employed, is not the composition for the highest possible ionic conductivity when utilizing Bi as an aliovalent dopant. However, for the processing parameters employed here, the tradeoff between composition and densification results in $Li_{6.25}La_3Zr_{1.25}Bi_{0.75}O_{12}$ to exhibit the highest ionic conductivity for a heat treatment of 900 °C for 10 hours. Alternative stoichiometries with even higher ionic conductivities may be achieved through the optimization of heat treatment and Bi doping.

The substitution of Bi into the LLZO garnet oxide structure through the Pechini method allows for a lower temperature formation of the desirable high ionic conductivity cubic phase. Bi additions also serve to decrease the temperature required for rapid grain growth and densification. As the content of Bi in the garnet is incremented, the degree of grain growth and densification is enhanced. Furthermore, ionic conductivity in $Li_{7-x}La_3Zr_{2-x}Bi_xO_{12}$ garnet type oxides is found to be a strong function of dopant amount. As such, garnets oxides may require optimized heat treatments for specific stoichiometries to attain the highest possible ionic conductivity.

FIGURES

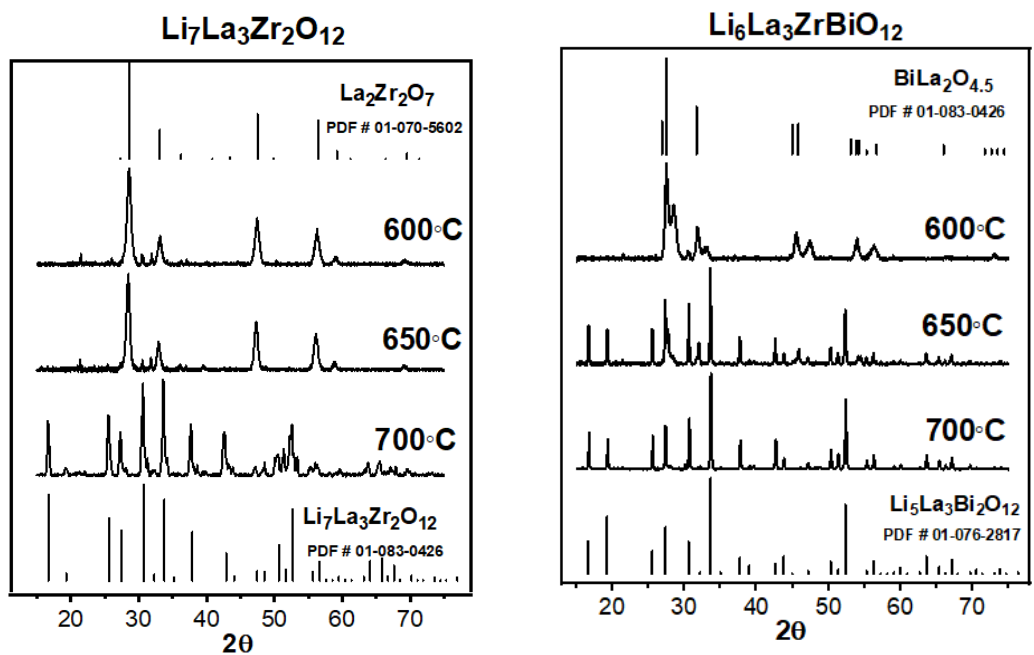

Fig. 1. X-ray diffraction spectra for (a) Li$_7$La$_3$Zr$_2$O$_{12}$ and (b) Li$_6$La$_3$ZrBiO$_{12}$ samples calcined at 600 °C, 650 °C, and 700 °C. Figure 2a shows patterns for un-doped LLZO garnets. These samples convert to the cubic phase at 700 °C, in agreement to similar studies[30]. In contrast, Figure 2b shows Bi-doped LLZO, which converts to cubic phase at 650 °C. Reference PDF spectra are provided to identify the crystal structures formed.

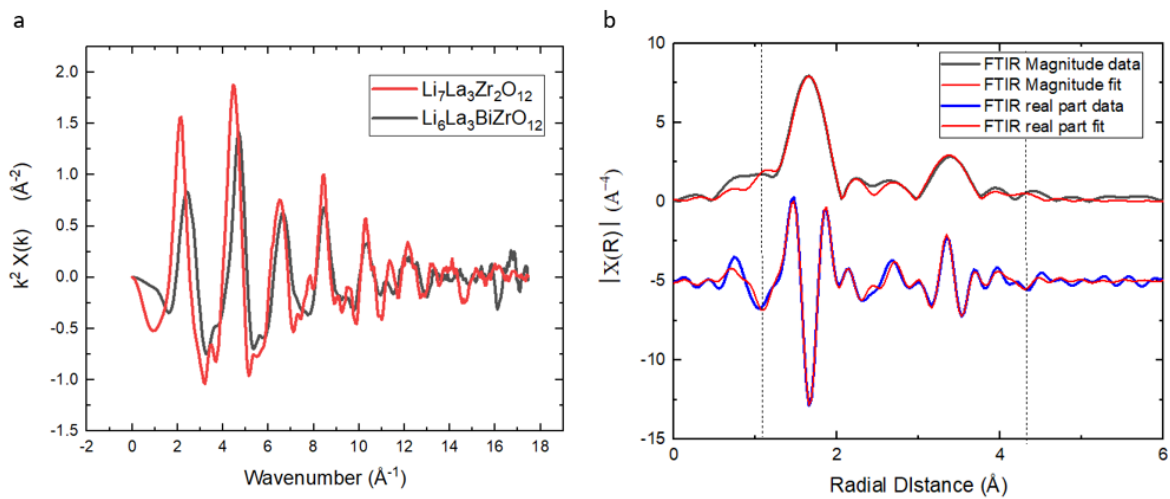

Fig. 2. (a) Representative EXAFS measurements at the Bi L$_3$-edge (black) and Zr K-edge (red) and (b) corresponding Fourier transform for the magnitude and real portion of the Bi XAF data and associated fitting using the Demeter software. Dotted vertical lines indicate the range covered by the fit.

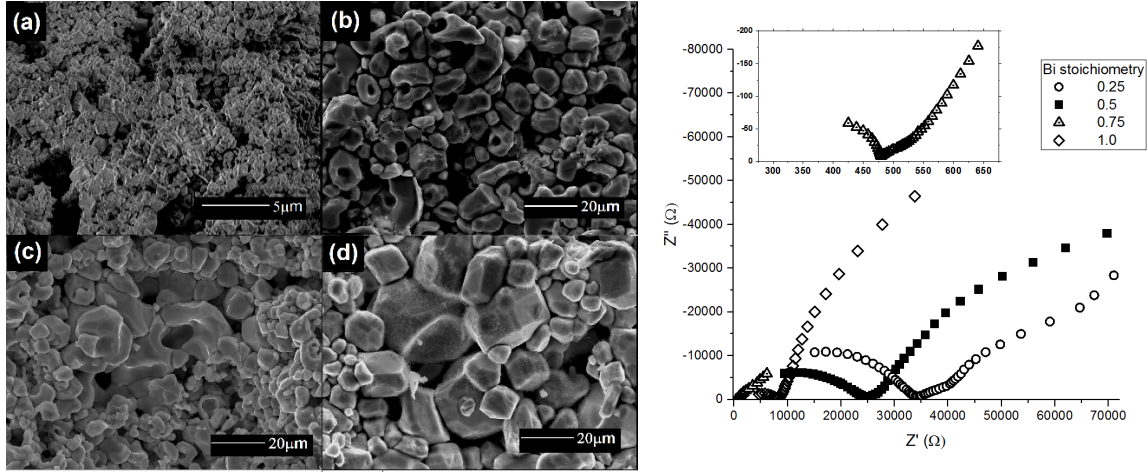

Fig 3. Doped LLZO microstructure and AC impedance measurements. (a-d) SEM images of pellet fracture surfaces for garnet oxides sintered at 900 °C for 10 hours. The sample stoichiometries are: (a) $Li_7La_3Zr_2O_{12}$, to (b) $Li_{6.5}La_3Zr_{1.5}Bi_{0.5}O_{12}$, to (c) $Li_{6.25}La_3Zr_{1.25}Bi_{0.75}O_{12}$, to (d) $Li_6La_3ZrBiO_{12}$. (e) Nyquist plots for pellet samples.

| Composition | Bi | Thickness (mm) | Density (g/cm$^3$) | Relative density | Ionic conductivity (S/cm) at 27 $^0$C |
|---|---|---|---|---|---|
| $Li_7La_3Zr_2O_{12}$ | 0 | 1.51 | 4.0 | 0.79 | Not measurable |
| $Li_{6.75}La_3Zr_{1.75}Bi_{0.25}O_{12}$ | 0.25 | 1.48 | 4.2 | 0.80 | 5.0 x 10$^{-6}$ |
| $Li_{6.5}La_3Zr_{1.5}Bi_{0.5}O_{12}$ | 0.5 | 1.30 | 4.4 | 0.81 | 7.2 x 10$^{-6}$ |
| $Li_{6.25}La_3Zr_{1.25}Bi_{0.75}O_{12}$ | 0.75 | 1.01 | 4.7 | 0.83 | 2.0 x 10$^{-4}$ |
| $Li_6La_3ZrBiO_{12}$ | 1.0 | 1.08 | 4.8 | 0.84 | 1.2 x 10$^{-5}$ |

Table I. Summary of pellet dimensions and physical properties for Bi-doped LLZO pellets sintered at 900 °C for 10 hours.

## SUPPLEMENTAL INFORMATION

**Synthesis**

Reagent grade chemicals of $LiNO_3$ (99.0% Sigma Aldrich), $La(NO_3)_3 \cdot 6H_2O$ (99.9% Alfa Aesar), $ZrO(NO_3)_2 \cdot xH_2O$ (99% Sigma Aldrich), and $Bi(NO_3)_3 \cdot 5H_2O$ (98% Alfa Aesar) were dissolved along with chelating agent citric acid into dilute nitric acid. After complete dissolution of the solids, ethylene glycol was added as a complexing agent of the polymerized mixture through polyesterfication of the chelated ionic compounds. To incorporate all the metal cations into the complex, a metallic ion to organic ratio of 38:62 was used. Additionally, to avoid auto-ignition of the resulting polymer upon pyrolysis, a citric acid to ethylene glycol ratio of 40:60 was used. The resulting solution was stirred at 70 °C until a thick transparent gel was formed. This gel was then heated at 120 °C to evaporate any remaining solvents, leaving behind a brown, rubbery solid. The resultant polymerized solid was analyzed via thermogravimetric analysis (TGA) to determine the decomposition temperature of the complex through heating in a platinum pan at 20 °C/min in air. In order to observe the formation of garnet oxides, the decomposition and subsequent oxidation of the polymerized-complex was characterized through TGA.

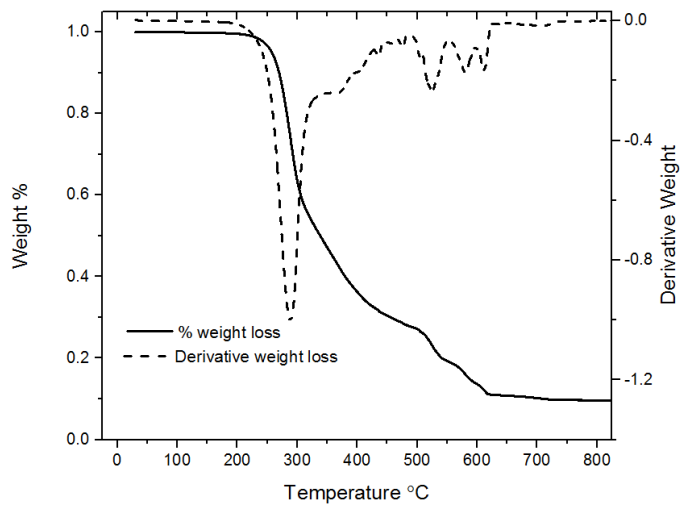

Fig. S1. TGA analysis exhibiting weight losses as a function of temperature for the synthesized complex. Mass loss at 300 °C results from the pyrolysis of the organic additives in the complex and from the oxidation of the remaining metallic ions to form the ceramic precursor powders.

Figure S1 displays weight loss as a function of temperature together with its derivative for the dried $Li_7La_3Zr_2O_{12}$ solid polymerized complex. Mass loss just below 300 °C is attributed to the rapid decomposition of the polymerized complex formed in the synthesis and dried in the oven. During this pyrolysis the metallic ions associated with the complex are oxidized to form the precursor oxide phases observed in the X-ray diffraction patterns. Gradual losses up to 650 °C are ascribed to the burn off of residual carbon. Small amounts of carbonate formation were observed in subsequent phase analysis, indicating incomplete oxidation of the metallic ions contained in the polymerized complex.

**Supplementary Table II**

| Sample | Excitation edge | Correlation | CN (set) | R (Å) | $\sigma^2$ ($10^{-3}$ Å$^2$) |
|---|---|---|---|---|---|
| $Li_7La_3Zr_2O_{12}$ | Zr-K | Zr-O | 6 | 2.108 ± 0.007 | 4.5 ± 0.7 |
| | | Zr-La | 6 | 3.636 ± 0.008[a] | 7.7 ± 0.7[b] |
| $Li_7La_3Zr_2O_{12}$ | La -L$_3$ | La-O | 8 | 2.585 ± 0.015 | 15.9 ± 1.8 |
| | | La-Zr | 4 | 3.636± 0.008[a] | 7.7 ± 0.7[b] |
| | | La-La | 4 | 3.986± 0.023 | 6.3 ± 2.6 |
| | | | | | |
| $Li_6La_3ZrBiO_{12}$ | Zr-K | Zr-O | 6 | 2.084 ± 0.012 | 7.8 ± 0.7 |
| | | Zr-La | 6 | 3.586 ± 0.022[c] | 11.9 ±1.6[d] |
| $Li_6La_3ZrBiO_{12}$ | La-L$_3$ | La-O | 8 | 2.581± 0.024 | 19.1 ± 2.3 |
| | | La-Bi | 2 | 3.656± 0.014[e] | 9.6 ± 1.2[f] |
| | | La-Zr | 2 | 3.586± 0.022[c] | 11.9 ± 1.6[d] |
| | | La-La | 4 | 4.020± 0.051 | 6.8 ± 3.9 |
| $Li_6La_3ZrBiO_{12}$ | Bi-L$_3$ | Bi-O | 6 | 2.097± 0.007 | 4.2 ± 0.4 |
| | | Bi-La | 6 | 3.656± 0.014[e] | 9.6 ± 1.2[f] |

Monochromatic x-rays were obtained using a fixed exit Si (111) monochromator. Harmonic rejection was accomplished using a rhodium coated mirror. Data were acquired at room temperature in transmission mode using gas ionization chambers as detectors. Samples were prepared as cold-pressed pellets using dry hexagonal boron nitride as the binder. Sample loading was adjusted to minimize artifacts from thickness effects.

Data reduction and analysis were performed using Athena and Artemis modules, implemented in the Demeter package [1]. The EXAFS spectra were then Fourier transformed to R-space and left uncorrected for photoelectron phase shifts. Structural parameters were obtained by fitting in R-space to ab initio theoretical paths generated using the FEFF6 code [2].

The Zr, La, and Bi data were co-refined for the Bi containing sample. Likewise, for the sample without Bi, the Zr and La data were co-refined. The coordination number (CN), distance (R), and the mean-squared relative displacement (a disorder related term, $\sigma^2$) of the dominant correlations in the fitted range are provided below. The coordination numbers were fixed at values expected for the garnet structure. Parameters with error bars were varied in the fits. A few structural parameters must be identical when evaluated from different vantage points; therefore, those parameters were constrained to take the same value (a-f). The many-body amplitude reduction factors ($S_0^2$) were fixed after preliminary refinements—0.98 (Zr), 0.76 (Bi), and 0.82 (La).

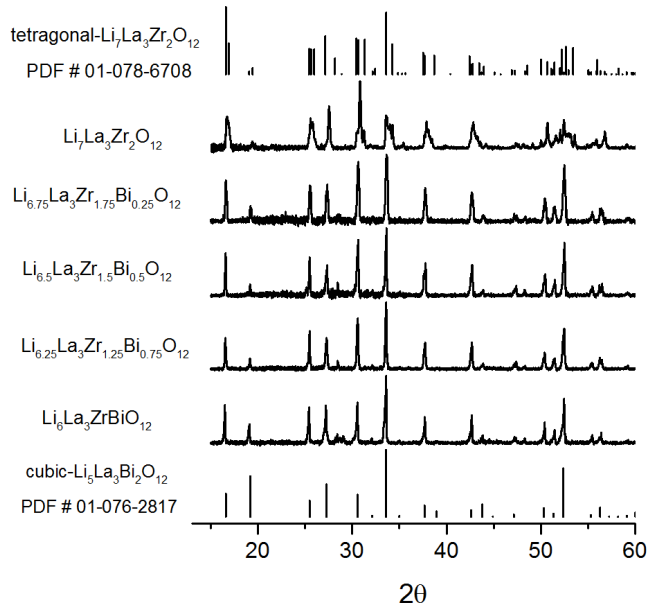

Fig. S2. X-ray diffraction patterns as a function of Bi composition for samples sintered at 900 °C for 10 hrs. Conversion to cubic garnet phases are seen for all samples containing Bi, whereas the un-doped LLZO sample exhibits a mixture of tetragonal and cubic phases. These diffraction patterns indicate that the cubic garnet phase in LiLaZrBiO is stable at the high temperatures required for densification. For comparison, this figure also provides the PDF patterns for tetragonal and cubic garnet phase.

Bi-doped LLZO pellets were calcined at 700 °C for 10 hours. The powders were grounded with agate mortar and pestle and pressed into 12 mm diameter pellets at 40 MPa for 10 minutes. They were covered in precursor powder to inhibit Li and Bi volatilization at the sintering temperature of 900 °C for 10 hours to create pellets approximately 10 mm in diameter and 1.0 mm thick. The corresponding diffraction patterns are compared in Fig. S2. The PDF reference diffraction patterns for both tetragonal and cubic garnet phases are also displayed. All samples containing any amount of Bi are stabilized as cubic Ia-3d garnet phases. In contrast, the $Li_7La_3Zr_2O_{12}$ sample stabilized as I41/acdZ tetragonal garnet phase - with a more ordered $Li^+$ sub-lattice. The stabilization of the cubic phase, even with a small amount of $Bi^{5+}$ indicates that small amounts of aliovalent dopants preferentially form this more disordered phase at 900 °C. As previously mentioned, this is likely caused by Bi-induced disorder in the $Li^+$ sub-lattice, thus providing energetically favorable conditions for the cubic structure to form over the more ordered tetragonal phase, which has closer to full $Li^+$ site occupancy. Furthermore, the higher amount of thermal energy allows for an un-doped $Li_7La_3Zr_2O_{12}$ sample to begin conversion from cubic phase observed in Fig. 2a at 700 °C to tetragonal phase at 900 °C. This increased thermal activation allows for additional ordering of the $Li^+$ ions in the overall structure and subsequent phase transformation to the more stable and ordered tetragonal garnet phase. However, such a conversion is not seen with any samples containing Bi due to the random distribution of $Li^+$ vacancies throughout the structure, which preferentially allows the phase to remain cubic.